\documentclass[11pt,letterpaper]{article}
\usepackage[utf8]{inputenc}
\usepackage[USenglish,spanish]{babel}
\usepackage{amsmath}
\usepackage{amsfonts}
\usepackage{amssymb}
\usepackage{graphicx}
\usepackage[left=3cm,right=3cm,top=2cm,bottom=3cm]{geometry}
\usepackage[small,bf]{caption} 

\author{Jorge Pinochet}

\title{\textbf{El límite de Chandrasekhar para principiantes\\ 
\large Chandrasekhar limit for beginners}}

\begin{document}

\author{Jorge Pinochet$^{*}$\\ \\
 \small{$^{*}$\textit{Departamento de Física, Universidad Metropolitana de Ciencias de la Educación,}}\\
 \small{\textit{Av. José Pedro Alessandri 774, Ñuñoa, Santiago, Chile.}}\\
 \small{e-mail: jorge.pinochet@umce.cl}\\}

\date{}
\maketitle

\begin{center}\rule{0.9\textwidth}{0.1mm} \end{center}
\begin{abstract}
\noindent En un breve artículo publicado en 1931 y ampliado en 1935, el astrofísico indio Subrahmanyan Chandrasekhar hizo público un importante descubrimiento astronómico donde introducía lo que en nuestros días se conoce como límite de Chandrasekhar. Este límite establece la masa máxima que puede alcanzar una enana blanca, que es el remanente estelar que se genera cuando una estrella de baja masa ha agotado su combustible nuclear. El presente trabajo tiene un doble propósito. El primero es presentar una derivación heurística del límite de Chandrasekhar. El segundo es clarificar la génesis del descubrimiento de Chandrasekhar, así como los aspectos conceptuales del tema. La exposición solo utiliza álgebra de secundaria, así como de algunas nociones generales de física clásica y teoría cuántica.\\

\noindent \textbf{Descriptores}: Límite de Chandrasekhar, enanas blancas, estudiantes no graduados de ciencias e ingeniería.   

\end{abstract}

\selectlanguage{USenglish}

\begin{abstract}
\noindent In a brief article published in 1931 and expanded in 1935, the Indian astrophysicist Subrahmanyan Chandrasekhar shared an important astronomical discovery where he introduced what is now known as Chandrasekhar limit. This limit establishes the maximum mass that a white dwarf can reach, which is the stellar remnant that is generated when a low mass star has used up its nuclear fuel. The present work has a double purpose. The first is to present a heuristic derivation of the Chandrasekhar limit. The second is to clarify the genesis of the discovery of Chandrasekhar, as well as the conceptual aspects of the subject. The exhibition only uses high school algebra, as well as some general notions of classical physics and quantum theory. \\ 



\begin{center}\rule{0.9\textwidth}{0.1mm} \end{center}
\end{abstract}

\selectlanguage{spanish}

\maketitle

\section{Introducción}
En un breve artículo publicado en 1931 y ampliado en 1935 [1, 2], el entonces joven astrofísico indio Subrahmanyan Chandrasekhar hizo público un notable descubrimiento que sorprendió a la comunidad astronómica. Los artículos contenían cálculos detallados que establecían la existencia de un límite superior para la masa que puede alcanzar una enana blanca\footnote{El término enana blanca fue introducido por el astrónomo americano Willem Jacob Luyten (1899-1994) en 1922. Sin embargo, al parecer el término fue popularizado más tarde por el astrofísico inglés Arthur Eddington (1882-1944).}, que es el remanente estelar que se genera cuando una estrella de baja masa ha agotado su combustible nuclear. Esta masa máxima para una enana blanca se conoce como \textit{límite de Chandrasekhar}, y su valor se estima en $1,4M_{\odot}$, donde $M_{\odot}$ es la masa solar [3, 4]. Aunque en sus artículos Chandrasekhar guardó silencio respecto del destino de un remanente estelar cuya masa supere el valor $1,4M_{\odot}$, su trabajo abrió las puertas para que otros abordaran esta importante interrogante, lo que más tarde condujo al descubrimiento de las estrellas de neutrones y los agujeros negros. Trascurridos casi 90 años desde el trabajo seminal del astrofísico indio, las observaciones astronómicas han confirmado el límite de Chandrasekhar, y las enanas blancas se han convertido en activo campo de investigación teórica y observacional.\\

Pero aparte de su innegable interés astronómico y científico, las enanas blancas ejercen un atractivo más personal y humano, ya que el destino que le espera al Sol cuando agote su combustible nuclear dentro de aproximadamente 5000 millones de años es convertirse en enana blanca, y este es también el destino de la inmensa mayoría de las estrellas, ya que el universo tiene una marcada preferencia por las estrellas con masas similares o menores que la solar.\\

El límite de Chandrasekhar es un tópico obligado en casi todos los textos generales de astronomía, así como en la mayoría de los cursos de astrofísica a nivel de pregrado. Típicamente, la derivación matemática del límite de Chandrasekhar requiere como mínimo el uso del cálculo diferencial e integral [3, 5-11], lo que limita el segmento de estudiantes que pueden acceder al tema. El presente trabajo tiene un doble propósito. El primero es presentar una derivación heurística del límite de Chandrasekhar, que solo utiliza álgebra de secundaria. El segundo es clarificar la génesis del descubrimiento de Chandrasekhar, así como los aspectos conceptuales del tema. \\

El artículo está organizado del siguiente modo. Primero se presenta una visión panorámica del tema, introduciendo las nociones de enana blanca y límite de Chandrasekhar. A continuación se analiza el principio de exclusión de Pauli y su relación con la presión de degeneración electrónica. Luego se estudia la configuración de equilibrio de una enana blanca en base a un cálculo no relativista aproximado, y se muestra que ese cálculo contradice el resultado de Chandrasekhar, discutiendo las razones de la discrepancia. Después se utiliza la teoría de la relatividad especial de Einstein para desarrollar un argumento más elaborado, pero también simple y directo, que permite obtener una formula aproximada para el límite de Chandrasekhar. Más adelante se analiza un importante descubrimiento astronómico que parece poner en cuestión el hallazgo del astrofísico indio. Para finalizar, se discute brevemente la importancia de la contribución científica de Chandrasekhar.
 
\section{El límite de Chandrasekhar: Una primera aproximación}

Una estrella es una enorme esfera de plasma incandescente [5, 9, 10]. Durante aproximadamente el 90\% de sus vidas, mientras las estrellas se encuentran en una etapa denominada \textit{secuencia principal}, prevalece un estado de equilibrio hidrostático, donde la fuerza de gravedad compresiva es contrarrestada por la presión térmica expansiva [5, 9, 10]. Esta última tiene su origen en el núcleo estelar, que es el lugar donde se producen las reacciones de fusión de hidrógeno en helio que generan la energía que hace brillar a las estrellas. Para sostenerse en el tiempo, la fusión termonuclear requiere temperaturas estables de $\sim10^{7} K$. Esta es, por lo tanto, la temperatura en el núcleo de las estrellas de secuencia principal.\\ 

Pero el combustible nuclear es finito, y cuando se agota, se detienen las reacciones termonucleares, lo que conlleva un descenso de la temperatura, y en definitiva, la muerte de la estrella. Haciendo abstracción de los complejos procesos físicos que se producen en esta etapa, el efecto global del cese de las reacciones nucleares es que la estrella se queda sin ninguna fuente de energía que luche contra el colapso gravitatorio. Como resultado, la estrella se contrae, reduciendo su radio y aumentando su densidad [5, 9, 10]. Cuando la densidad se vuelve muy grande, aparece un tipo de presión de origen cuántico denominada \textit{presión de degeneración}, que se opone al colapso. En rigor, la presión de degeneración siempre está presente, pero durante la secuencia principal esta presión es tan pequeña que puede ignorarse. Sin embargo, después del cese de las reacciones nucleares, la presión de degeneración se hace muy grande, y frente a ella, la presión térmica se vuelve despreciable [5, 9, 10]. \\ 

Los objetos que se forman como resultado del agotamiento del combustible nuclear de las estrellas se denominan \textit{remanentes estelares}, y las enanas blancas son el remanente de estrellas de baja masa, es decir, estrellas que durante la etapa en la secuencia principal tienen masas menores que $\sim 8 M_{\odot}$ [3]. Como se discutirá con más detalle en las siguientes secciones, en las enanas blancas, la presión de degeneración se manifiesta como una repulsión entre electrones que se opone al colapso gravitatorio. De hecho, a las densidades observadas en las enanas blancas, que son de $\sim (10^{3} - 10^{4}) kg \cdot cm^{-3}$, la presión de degeneración está asociada principalmente a los electrones, ya que la presión asociada a los protones y neutrones es muy baja y puede despreciarse [5, 9, 10]. En una enana blanca, la presión de degeneración electrónica contrarresta exactamente a la atracción gravitacional, de manera que prevalece un estado de perfecto equilibrio [3]. Este es el destino que le espera al Sol cuando agote su combustible nuclear dentro de $\sim 5\times 10^{9}$ \textit{años}. \\ 

La presión de degeneración electrónica es un fenómeno radicalmente diferente de la presión térmica, que es la que mantiene en equilibrio a las estrellas de secuencia principal, como es el caso del Sol. Un aspecto importante de la presión de degeneración es que es independiente de la temperatura [6]. En teoría, cuando la temperatura global se aproxima a $0 K$, una enana blanca debería convertirse en una enana negra, vale decir, una esfera fría e invisible de materia densa. Sin embargo, el universo no es lo suficientemente viejo para albergar enanas negras, ya que el tiempo requerido para el enfriamiento de una enana blanca es del orden de la edad del universo ($\sim 10^{10}$ \textit{años}). \\ 

¿La presión de degeneración electrónica siempre puede detener el colapso gravitatorio? En otras palabras, ¿el destino final de toda estrella moribunda es convertirse en una enana blanca? El gran hallazgo de Chandrasekhar consistió en demostrar que si se combinan la teoría de la gravitación de Newton con la mecánica cuántica y la teoría de la relatividad especial de Einstein, se encuentra que la presión de degeneración electrónica es capaz de sostener el colapso si y solamente si la masa del remanente estelar no supera un valor dado por:

\begin{equation}
M_{Ch} =3.14\left(\dfrac{Z}{A} \right)^{2} \left( \frac{\hbar c}{G m_{p}^{2}}\right)^{3/2} m_{p}.   
\end{equation}

Esta es la fórmula que establece la \textit{masa límite de Chandrasekhar} $M_{Ch}$, más conocida como  \textit{límite de Chandrasekhar} [3, 10], donde $Z$ es el número atómico de los núcleos que componen la enana blanca (número de protones), $A$ es el número de masa (número de protones y de neutrones), $m_{p} = 1,67 \times 10^{-27} kg$ es la masa del protón, $\hbar = h/2\pi = 1,05 \times 10^{-34} J\cdot s$ es la constante de Planck reducida, $G = 6,67\times 10^{-11} N\cdot m^{2} \cdot kg^{-2}$ es la constante de gravitación universal y $c = 3\times 10^{8} m\cdot s^{-1}$ la rapidez de la luz en el vacío. Una enana blanca típica está formada por Helio, Carbono y Oxígeno, de modo que $Z/A \approx 0.5$. Si se introducen las cifras anteriores en la Ec. (1) se encuentra que $M_{Ch} \approx 1,4M_{\odot}$. Luego, una enana blanca no puede tener una masa superior a esta cifra.

\section{El Principio de exclusión y la presión de degeneración}

Las partículas fundamentales tienen una propiedad denominada \textit{spin} que corresponde a su momentum angular intrínseco, y que se mide en términos de la constante de Planck reducida $\hbar = h/2\pi$. Las partículas elementales con valores enteros de spin ($0\hbar, 1\hbar, 2\hbar$, etc.) se denominan bosones, mientas que las partículas con valores semienteros ($\hbar/2, 3\hbar/2, 5\hbar/2$, etc.) se llaman fermiones. El principio de exclusión solo se aplica a los fermiones, que incluyen al protón, al neutrón y al electrón, cuyos valores posibles de spin tienen magnitud $\hbar/2$. Aplicado a esta clase de partículas, el principio de exclusión establece que una determinada región muy pequeña del espacio puede contener un máximo de dos electrones con la misma energía al mismo tiempo, uno con spin $+ \hbar/2$ y el otro con spin $- \hbar/2$ [12, 13]. \\

Por lo tanto, el principio de exclusión se manifiesta como una repulsión entre fermiones idénticos cuando son obligados a reducir el volumen de espacio donde están confinados. Dicha repulsión puede interpretarse como una presión expansiva denominada \textit{presión de degeneración} [13]. Como se ha mencionado antes, la presión de degeneración es radicalmente diferente de la presión térmica que mantiene en equilibrio a las estrellas de secuencia principal, ya que la presión de degeneración es independiente de la temperatura, y depende únicamente de la densidad.  \\ 

Mediante el principio de indeterminación de Heisenberg es posible obtener una relación matemática aproximada para el volumen mínimo que puede ocupar un conjunto de fermiones idénticos sometidos a una fuerza compresiva, como la generada por la contracción gravitatoria de un remanente estelar [14]. Una de las formas que puede adoptar este principio es la siguiente [12, 13]: 

\begin{equation}
\Delta p \Delta x \geq \frac{\hbar}{2},
\end{equation}

donde $\Delta p$ es la indeterminación en el momentum lineal de una partícula y $\Delta x$ es la indeterminación en la posición. Para un valor fijo de $\Delta p$, la celda de volumen más pequeña compatible con el principio de indeterminación es del orden de $ \Delta x^{3} \approx \hbar^{3} / (\Delta p)^{3}$. Consideremos un volumen $V$ compuesto por un gran número de celdas de tamaño $\Delta x^{3}$, y supongamos que $V$ contiene $N_{f}$ fermiones idénticos. De acuerdo con el principio de exclusión, $V$ será mínimo cuando cada celda contenga un máximo de dos fermiones idénticos, uno con spin $+ \hbar/2$ y otro con spin $- \hbar/2$. Por lo tanto, el valor mínimo de $V$ será [3, 14]:

\begin{equation}
V_{min} \sim \frac{1}{2} N_{f} \Delta x^{3}.
\end{equation}

El estado físico en el que se encuentran los fermiones confinados en $V_{min}$ se denomina materia degenerada, y es el estado que adoptan los electrones en el interior de una enana blanca.

\section{Enanas blancas no relativistas: El enfoque de Fowler}

En la medida que consideramos enanas blancas más masivas, la gravedad que estos objetos ejercen sobre sí mismos es mayor, provocando que el volumen $\Delta x^{3}$ de las celdas donde están confinados los electrones disminuya. De acuerdo con el principio de incertidumbre. Ec. (2), lo anterior implica que la incerteza en el momentum lineal $\Delta p$ debe aumentar; esto conlleva un incremento en la rapidez y en la energía cinética media de los electrones, lo que les permite contrarrestar la fuerza gravitacional que la enana blanca ejerce sobre sí misma. El primero en llegar a esta importante conclusión fue el astrofísico inglés Ralph Fowler en un artículo publicado en 1926, donde por primera vez se aplicaba la flamante mecánica cuántica al análisis de las condiciones de equilibrio en una enana blanca [15]. El resultado fundamental que se deriva de los cálculos de Fowler es la denominada \textit{relación masa-radio} para una enana blanca. Según esta relación, en la medida que aumenta la masa, aumenta la fuerza contractiva de la gravedad, y por tanto disminuye el radio de la enana blanca.\\  

Para comprender los alcances y los límites del hallazgo de Fowler, a continuación se desarrolla un sencillo argumento heurístico basado en un modelo semiclásico simplificado del interior de un remanente estelar. Pese a estas simplificaciones, el modelo proporciona excelentes resultados, tal como quedará en evidencia en ésta y en la próxima sección. El modelo se basa en el concepto de energía, y por lo tanto no hace uso explícito de la presión de degeneración. En otro trabajo, el autor ha elaborado una derivación heurística del límite de Chandrasekhar que hace uso directo de la presión de degeneración electrónica [14]. \\ 

Como en las derivaciones heurísticas las constantes adimensionales son poco confiables, por simplicidad las omitiremos de los cálculos. Al igual que en los trabajos originales de Chandrasekhar, en adelante supondremos que la enana blanca carece de rotación.\\ 

Consideremos un remanente estelar de masa $M$ y volumen $V_{min}$, compuesto por $N_{e}$ electrones. Definimos el número $N$ de electrones por unidad de masa como:

\begin{equation}
N \equiv \frac{N_{e}}{M}.
\end{equation}

Tomando $N_{e} = N_{f}$ en la Ec. (3) se tiene:

\begin{equation}
\Delta x^{3} \sim \frac{V_{min}}{N_{e}}.
\end{equation}

Por otra parte, la energía potencial total $U$ del remanente estelar de radio $R$ es:

\begin{equation}
U \sim - \frac{GM^{2}}{R}.
\end{equation}

Debido a la gran rapidez de los electrones dentro de la enana blanca, se puede demostrar que se comportan como partículas libres, es decir, no interactúan entre sí, de modo que su energía total será puramente cinética. Si suponemos que en promedio todos los electrones poseen la misma energía cinética, la energía cinética total vendrá dada por: 

\begin{equation}
K \sim N_{e} \frac{p^{2}}{m_{e}},
\end{equation}

donde $m_{e} = 9,1 \times 10^{-31} kg$ es la masa del electrón. Empleando el principio de indeterminación, Ec. (2), e introduciendo la Ec. (5):

\begin{equation}
p \sim \Delta p \sim \frac{\hbar}{\Delta x} \sim \hbar \left( \frac{N_{e}}{V_{min}} \right) ^{1/3}. 
\end{equation}

Así, partir de la Ec. (8), la Ec. (7) queda\footnote{El paso de la Ec. (8) a la (9) amerita un comentario. Notemos que la Ec. (8) corresponde al momentum lineal de un electrón individual, pero al pasar a la Ec. (9), sumamos sobre el momentum de un número de electrones $N_{e}$ muy grande. Por otra parte, sabemos que el momentum lineal es una cantidad vectorial, cuya suma sobre un rango amplio de valores aleatorios es nulo, ya que se suman las contribuciones en direcciones opuestas. No obstante, los valores calculados en la Ec. (9) no son nulos. La explicación a esta aparente discrepancia es simple: En la Ec. (9) se calcula la suma sobre el momentum al cuadrado, lo que implica que todos los valores son positivos, y por lo tanto no se anulan. Como muestra la Ec. (7), esto se debe a que la energía cinética, que es una cantidad escalar, depende del cuadrado del momentum lineal. De hecho, no podría ser de otro modo, pues de lo contrario, la energía cinética tendría direccionalidad y no sería un escalar.}:

\begin{equation}
K \sim \frac{N_{e}}{m_{e}} \hbar^{2} \left( \frac{N_{e}}{V_{min}} \right) ^{2/3}.
\end{equation}

Usando el hecho que $V$ es proporcional a $R^{3}$, e introduciendo la Ec. (4) en la (9):

\begin{equation}
K \sim \frac{NM \hbar^{2}}{m_{e}} \left( \frac{NM}{V_{min}} \right) ^{2/3} = \frac{\hbar^{2} N^{5/3} M^{5/3}}{m_{e} R^{2}}.
\end{equation}

La energía total $E$ del remanente estelar vendrá dada por $K + U$, de modo que:

\begin{equation}
E \sim \frac{\hbar^{2} N^{5/3} M^{5/3}}{m_{e} R^{2}} - \frac{GM^{2}}{R}.
\end{equation}

La condición para que exista equilibrio entre la presión de degeneración y la compresión gravitacional es que $E = K + U \approx 0$. En efecto, si $K + U \gg 0$, entonces $R$ debe aumentar y el núcleo estelar no puede convertirse en una enana blanca. Por el contrario, si $K + U \ll 0$, entonces $R$ debe disminuir, y la presión de degeneración no puede detener el colapso. Tomando $E \approx 0$ en la Ec. (11) se obtiene: 

\begin{equation}
\frac{\hbar^{2} N^{5/3} M^{5/3}}{m_{e} R^{2}} \approx \frac{GM^{2}}{R}.
\end{equation}

Resolviendo para $R$:

\begin{equation}
R(M) \sim \frac{N^{5/3} \hbar^{2}}{m_{e} G} M^{-1/3}.
\end{equation}

Cálculos más detallados muestran que [3]:

\begin{equation}
R(M) \approx 1,97 \frac{N^{5/3} \hbar^{2}}{m_{e} G} M^{-1/3}.
\end{equation}

Las Ecs. (13) y (14) expresan la relación masa-radio para una enana blanca. Tal como habíamos anticipado, estas ecuaciones muestran que la masa del remanente estelar es inversamente proporcional a su radio (al cubo). Por lo tanto, para cualquier valor de $M$ siempre existirá un valor de $R$ para el cual la presión de degeneración pueda detener el colapso gravitatorio (ver Fig. 1). En otras palabras, las Ecs. (13) y (14) sugieren que toda estrella terminará su vida como enana blanca. Estas ecuaciones se encuentran en flagrante conflicto con la Ec. (1) obtenida por Chandrasekhar. La explicación para esta discrepancia se debe a que los cálculos realizados en esta sección no tienen en cuenta que de acuerdo con la teoría de la relatividad especial de Einstein, la velocidad de los electrones no puede superar la velocidad de la luz en el vacío. Sin embargo, al emplear la expresión clásica para la energía cinética, dada por la Ec. (7), se ha supuesto implícitamente que la velocidad de los electrones puede aumentar sin límite.\\

\begin{figure}
  \centering
    \includegraphics[width=0.9\textwidth]{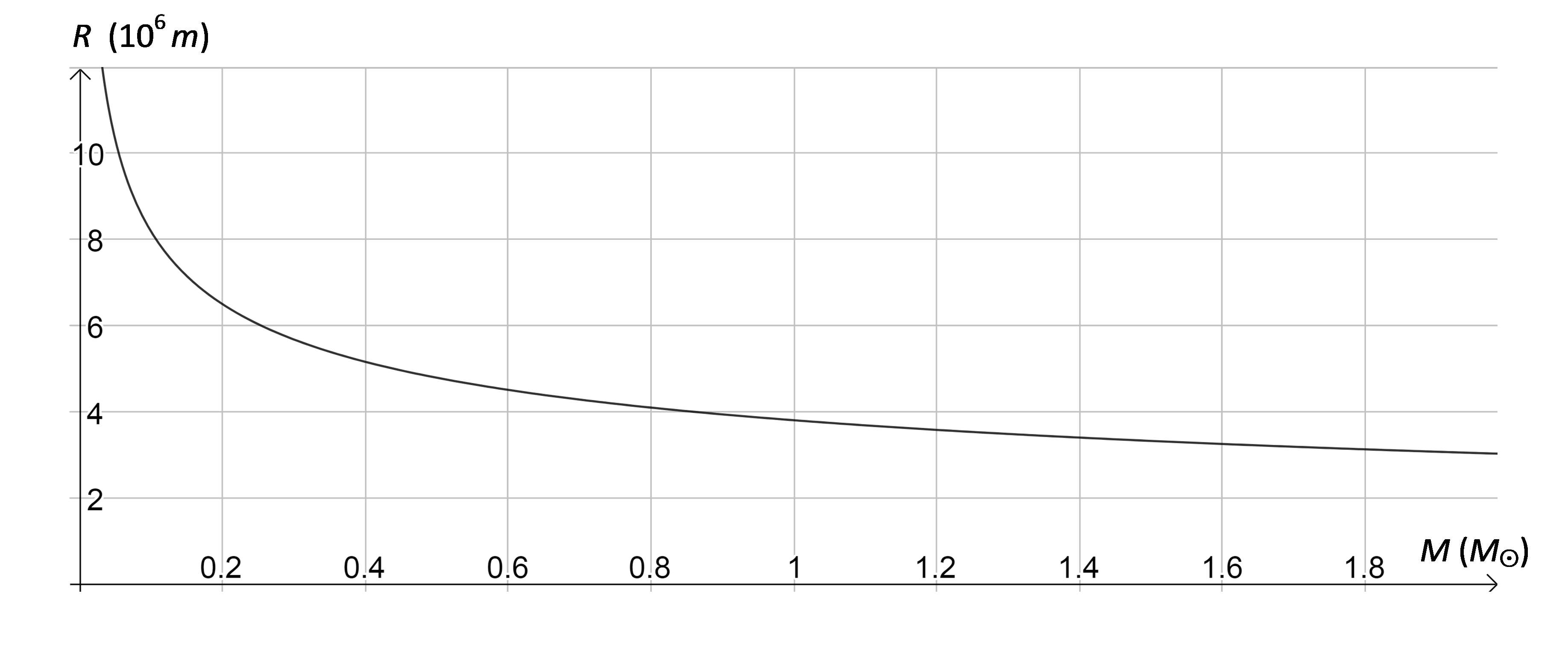}
  \caption{Relación masa-radio ($M-R$) para una enana blanca, según los cálculos de Fowler (la ecuación graficada se presenta en la siguiente sección). El eje $R$ está expresado en unidades de $10^{6} m$, y el eje $M$ está en unidades de masa solar, $M_{\odot}$.}
\end{figure}

De este modo, sin importar que tan grande sea la masa del remanente estelar, ni que tan poderosa sea la fuerza gravitacional generada por el remanente sobre si mismo, los electrones siempre podrán moverse con una rapidez lo suficientemente elevada para detener el colapso gravitatorio y conducir a una configuración final de equilibrio. En la siguiente sección se analizará esta situación en detalle, donde mostraremos que para llegar a la conclusión de que existe un límite superior para la masa de una enana blanca, que fue el gran hallazgo de Chandrasekhar, es necesario incorporar en los cálculos la teoría de la relatividad especial.

\section{Enanas blancas relativistas y el límite de Chandrasekhar}

Recordemos que en la medida que la masa de una enana blanca aumenta, el volumen $\Delta x^{3}$ de las celdas donde están confinados los electrones disminuye, y de acuerdo con el principio de incertidumbre. Ec. (2), esto implica que la rapidez y la energía cinética media de los electrones debe crecer. Si la energía cinética se hace lo suficientemente elevada, la rapidez de los electrones se aproxima a la de la luz, lo que implica que deben tomarse en cuenta los efectos de la relatividad especial. Esta fue la conclusión a la que llegó Chandrasekhar cuando revisó los cálculos de Fowler, y se percató de que el astrofísico inglés no había considerado la teoría de Einstein. Esto significa que los cálculos no relativistas de Fowler solo son una aproximación válida cuando la masa de una enana blanca es pequeña. Así, Chandrasekhar emprendió la compleja tarea de reformular y extender los cálculos de Fowler para el caso de enanas blancas masivas y densas, incorporando la teoría de Einstein, lo que le condujo a descubrir el límite que lleva su nombre [1, 2]. \\ 

A continuación se desarrolla un sencillo argumento heurístico que permite obtener una fórmula aproximada para el límite de Chandrasekhar. Para ello, vamos a utilizar nuevamente el modelo simplificado de la sección anterior, basado en el concepto de energía. Por lo tanto, el modelo no hace uno explícito de la presión de degeneración. Por simplicidad, continuaremos suponiendo que la enana blanca carece de rotación. 
\\ 

Consideremos una enana blanca lo suficientemente masiva y densa para que la velocidad media de los electrones confinados dentro de ella sea cercana a la rapidez de la luz en el vacio, $c$. Nuevamente, debido a la gran rapidez de los electrones, podemos asumir que se comportan como partículas libres. Bajo estas condiciones, la Ec. (7), que corresponde a la fórmula clásica para la energía total de una partícula libre, no es aplicable. En su reemplazo debemos utilizar la expresión relativista para la energía total $E$ de una partícula libre [13]:

\begin{equation}
E = \sqrt{(pc)^{2} + (m_{e} c^{2})^{2}} = pc \sqrt{1+\left( \frac{m_{e} c}{p}\right)^{2} },
\end{equation}

donde $p$ es el momentum lineal relativista del electrón, y $m_{e}$ es su masa en reposo, que es constante. Es importante recordar que esta fórmula, al igual que todas las ecuaciones de la relatividad especial, se basa en el postulado de que nada en el universo puede moverse más rápido que $c$. Esto marca una diferenca fundamental con la mecánica clásica, donde no existe un límte superior para la rapidez que puede tener un cuerpo o una señal lúminosa. \\ 
 
Regresando a la Ec. (15), como los electrones dentro de la enana blanca masiva se mueven con gran rapidez, el término constante $m_{e} c$ es despreciable frente al término variable $p$ que depende de la denominada \textit{masa relativista}\footnote{En el caso unidimensional, el momentum lineal relativista $p$ de una partícula de masa en reposo $m_{0}$ y rapidez $v$ se define como $p = mv$ donde $m = m_{0}/\sqrt{1-v^{2}/c^{2}}$ es la masa relativista. Se observa que cuando $v\rightarrow c$, $m \rightarrow \infty$.}, que crece sin límite cuando la rapidez de la partícula se aproxima a $c$, de manera que $m_{e} c / p \approx 0$ y la energía total del electrón $E$ es puramente cinética:

\begin{equation}
E \approx K \approx pc.
\end{equation}

Si suponemos que en promedio todos los electrones poseen la misma energía cinética, entonces la energía cinética relativista total de los electrones dentro de la enana blanca vendrá dada aproximadamente por $N_{e} pc$. Notemos que esta expresión difiere de la relación clásica empleada en la sección anterior, Ec. (7). Utilizando la expresión para el momentum $p$ dada por la Ec. (8), la energía cinética relativista total será:

\begin{equation}
K \sim N_{e} pc \sim N_{e} \hbar c \left( \frac{N_{e}}{V_{min}} \right) ^{1/3}.
\end{equation}

En forma análoga a como se procedió en la sección anterior, como $V_{min}$ es proporcional a $R^{3}$, a partir de la Ec. (4) la Ec. (17) puede escribirse como:

\begin{equation}
K \sim NM \hbar c \left( \frac{NM}{R^{3}} \right) ^{1/3} = \frac{N^{4/3} \hbar c M^{4/3}}{R}.
\end{equation}

Usando el mismo argumento de la Sección 4, la condición para que exista equilibrio entre la presión de degeneración y la compresión gravitacional es que la energía total sea nula. A partir de las Ecs. (6) y (18), tomando $K + U \approx 0$ resulta:

\begin{equation}
\frac{N^{4/3} \hbar c M^{4/3}}{R} \approx \frac{GM^{2}}{R}.
\end{equation}

En este punto del razonamiento sucede algo que resulta trivial desde un punto de vista matemático, pero que tiene una profunda significación física. A diferencia de lo que ocurrió en la sección anterior con la Ec. (12), en la Ec. (19) $R$ se cancela, y por lo tanto obtenemos un resultado que es independiente del radio de la enana blanca. Esta cancelación es consecuencia directa de haber incorporado en los cálculos la relatividad especial. Así, a partir de la Ec. (19) obtenemos un valor para $M$ que es independiente de $R$ y que por lo tanto es constante:

\begin{equation}
M \sim N^{2} \left( \frac{\hbar c}{G} \right) ^{3/2}.
\end{equation}

Esta es la versión aproximada del límite de Chandrasekhar que nos propusimos derivar, pero para que esta expresión sea equivalente a la Ec. (1) debemos determinar el valor de $N$. Para ello, comencemos recordando que un núcleo atómico está compuesto por $A$ nucleones y $Z$ protones. La masa de un nucleón es $\sim 2000$ veces mayor que la de un electrón. Si se desprecia la masa de los electrones, el número de núcleos $N_{nuc}$ contenidos en el volumen $V_{min}$ del remanente estelar será igual al cociente entre $M$ y la masa de un núcleo. Como la masa de un nucleón es del orden de la masa de un protón $m_{p}$, la masa de cada núcleo será del orden de $Am_{p}$, de modo que $N_{nuc} = M/Am_{p}$. Si hay $Z$ protones por núcleo, el número total de protones será $ZN_{nuc} = ZM/Am_{p}$. Si se asume que el remanente estelar es eléctricamente neutro, el número de electrones $N_{e}$ debe ser igual al de protones [14]:

\begin{equation}
N_{e} = \frac{ZM}{Am_{p}}.
\end{equation}

Combinando las Ecs. (4) y (20):

\begin{equation}
N = \frac{Z}{Am_{p}}.
\end{equation}

Introduciendo la Ec. (22) en la (20) obtenemos finalmente una expresión que es formalmente idéntica a la Ec. (1) para el límite de Chandrasekhar:

\begin{equation}
M \sim \left( \frac{Z}{Am_{p}} \right) ^{2} \left( \frac{\hbar c}{G} \right) ^{3/2} = \left(\dfrac{Z}{A} \right)^{2} \left( \frac{\hbar c}{G m_{p}^{2}}\right)^{3/2} m_{p}.
\end{equation}

Se observa que esta expresión difiere en un factor 3,15 de la Ec. (1) para $M_{Ch}$. Es importante enfatizar que a diferencia de las Ecs. (13) y (14), obtenidas empleando la expresión clásica para la energía cinética de los electrones, la Ec. (23) señala que existe una masa límite más allá de la cual la presión de degeneración electrónica no es capaz de detener el colapso gravitatorio. \\ 

Desde el punto de vista de la relatividad especial, la interpretación de la Ec. (23) es la siguiente. Imaginemos que disponemos de un mecanismo para inyectar masa a una enana blanca. En la medida que aumentamos $M$, también aumentamos la fuerza gravitacional, lo que conlleva una disminución del radio. De acuerdo con el principio de incertidumbre, una reducción en el radio provoca una disminución del volumen disponible para el movimiento de los electrones, lo que genera que éstos aumenten su momentum lineal y su rapidez. Sin embargo, llegará un momento en que la rapidez se aproxime a $c$. Desde ese instante los electrones no podrán continuar incrementando su rapidez, aunque en principio nada impide continuar inyectando masa a la estrella indefinidamente. Se concluye entonces que una vez que los electrones han alcanzado una rapidez muy cercana a $c$, cualquier aumento de $M$ conducirá a una contracción gravitacional que no podrá ser detenida por la presión de degeneración electrónica. Así, el valor de $M_{Ch}$ es independiente del radio de la enana blanca. \\ 

En su artículo de 1935, Chandrasekhar efectuó cálculos numéricos detallados, repetidos después por otros especialistas, que le permitieron encontrar la relación masa-radio para una enana blanca relativista [2]. Es decir, Chandrasekhar obtuvo una gráfica $R(M)$. Para comparar el modelo relativista de Chandrasekhar con el modelo no relativista de Fowler, podemos reescribir la Ec. (14) tomando el valor de $N$ dado por la Ec. (22), e introduciendo los valores de las constantes numéricas. Al proceder de este modo, y tomando $Z/A \approx 0,5$, tal como hicimos en la Sección 2 para obtener el valor $M_{Ch} \approx 1,4M_{\odot}$, resulta [3]: 

\begin{equation}
R(M) \approx 3,8 \times 10^{6} m \left( \frac{M}{M_{\odot}} \right) ^{-1/3},
\end{equation}

donde $M_{\odot} = 1,99 \times 10^{30} kg$ es la masa solar. La Fig. 2 muestra el modelo relativista (en color verde), el modelo no relativista (en negro) dado por la Ec. (24), y el límite de Chandrasekhar (en rojo). Se observa que para valores de $M$ pequeños ($M/M_{\odot} \ll 1$), y por tanto valores de $R$ comparativamente grandes, las predicciones concuerdan, pero divergen gradualmente en la medida que $M \rightarrow M_{Ch} \approx 1,4M_{\odot}$. También se aprecia que según el modelo de Chandrasekhar, si $M$ aumenta, la gravedad crece y por lo tanto $R$ disminuye. Hasta este punto, todo concuerda con lo que hemos discutido antes. Sin embargo, el modelo relativista también predice que $R \rightarrow 0$ cuando $M \rightarrow M_{Ch}$, lo que claramente es poco realista.\\  

Para interpretar correctamente este último resultado es necesario recordar los supuestos simplificadores en los que se basa el modelo de Chandrasekhar. Por una parte, el modelo supone que la presión de degeneración es producida únicamente por los electrones. Por otra parte supone que los electrones son libres, lo que significa que no interactúan entre si. Un sistema ideal de electrones libres se denomina \textit{gas de Fermi}, y una de sus propiedades es que el volumen del sistema, y por tanto su radio (suponiendo forma esférica), puede tender a cero. En el modelo de Chandrasekhar se asume que las enanas blancas son esferas de gas ideal de Fermi. Luego, en el momento en que la velocidad de los electrones se aproxima a $c$ y por tanto no puede seguir amentando, la presión de degeneración es incapaz de detener el colapso y la gravedad ya no tiene contrapeso. Por lo tanto, cuando $M_{Ch} \approx 1,4M_{\odot}$ la esfera ideal de gas de Fermi solo puede responder reduciendo su volumen y su radio a cero. Evidentemente, bajo condiciones más realistas, el radio de una enana blanca no puede hacerse cero. \\

\begin{figure}
  \centering
    \includegraphics[width=0.9\textwidth]{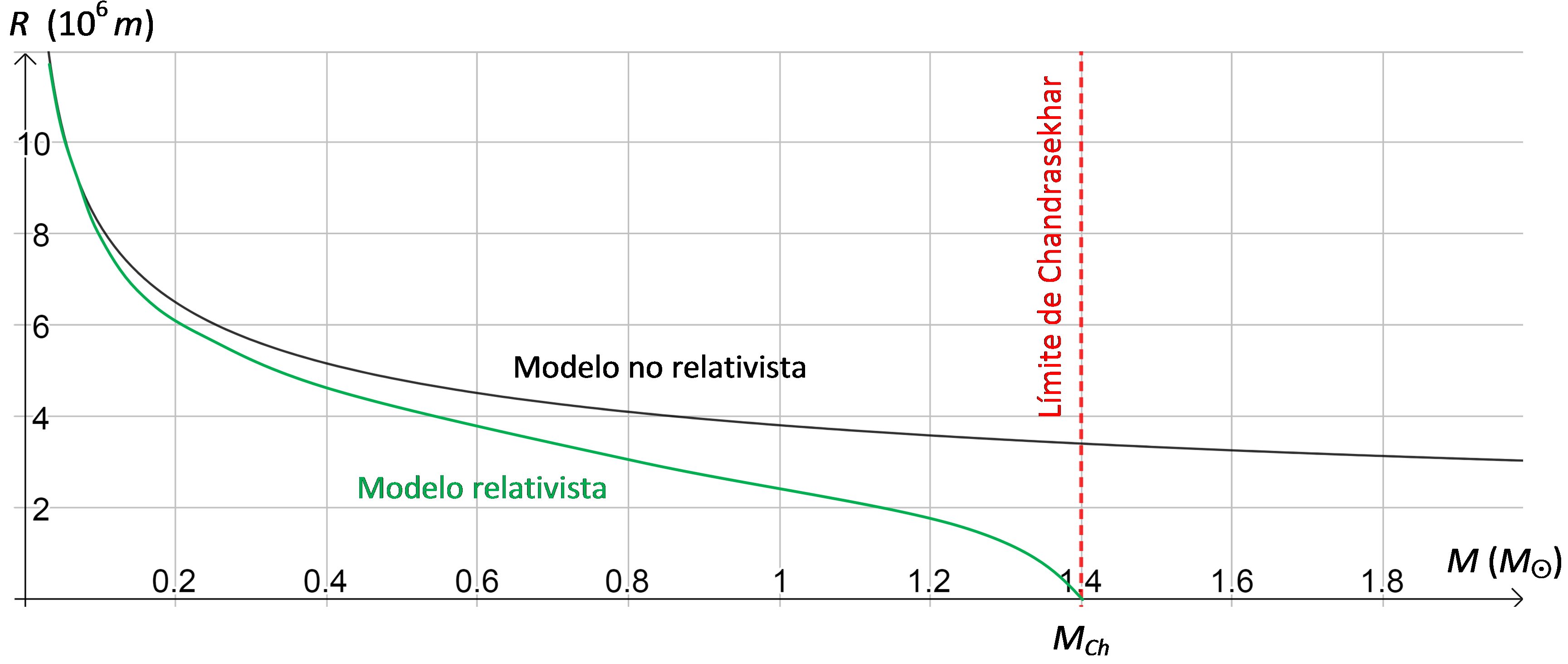}
  \caption{Modelo relativista (verde) y no relativista (negro). Para $M_{\odot}/M \ll 1$, las predicciones son muy similares, pero divergen en la medida que $M$ se aproxima a $1,4M_{\odot}$. La curva verde es solo una representación cualitativa de la relación masa-radio para el caso relativista.}
\end{figure}

¿Qué sucede con un remanente estelar cuya masa $M$ supera el límite de Chandrasekhar? Como se ha señalado en la introducción, aunque en sus artículos de 1931 y 1935 Chandrasekhar no respondió esta pregunta, su trabajo abrió las puertas para que otros abordaran esta importante interrogante. Lo que sabemos hoy en día gracias a décadas de trabajo de investigación, es que una vez que un remanente supera el valor $M_{Ch} \approx 1,4M_{\odot}$ solo existen dos posibilidades:\\ 

\textbf{Posibilidad 1}. Cuando $\sim 1,4M_{\odot} < M < \sim 3M_{\odot}$ el colapso gravitatorio provoca que los electrones y los protones comiencen a fusionarse para formar neutrones, hasta que el remanente estelar alcanza una configuración de equilibrio conocida como \textit{estrella de neutrones}. A diferencia de lo que ocurre con una enana blanca, en una estrella de neutrones el colapso es detenido por una nueva manifestación de la presión de degeneración, que ahora se debe a la repulsión entre neutrones degenerados [6]. En una estrella de neutrones la densidad alcanza valores de $\sim 10^{11} kg \cdot cm^{-3}$, una cifra que equivale a aproximadamente $10^{7}$ veces la densidad media de una enana blanca. El valor máximo para la masa de una estrella de neutrones, $M \sim 3M_{\odot}$, se conoce como \textit{límite de Tolman-Oppenheimer-Volkoff} en honor de los físicos que lo calcularon por primera vez [16, 17], y está sujeto a incertidumbre porque aún no se conoce bien la ecuación de estado de la materia hadrónica a alta densidad\footnote{Recordemos que los protones y neutrones son hadrones, nombre con el que se designa a las partículas subatómicas compuestas por quarks, y que permanecen unidas por la interacción fuerte entre ellas.}  [10, 18]. \\ 

\textbf{Posibilidad 2}. Cuando $M > \sim 3M_{\odot}$ no se conoce fuerza en la naturaleza capaz de detener el colapso gravitatorio. En ese momento entra en escena uno de los objetos astronómicos más enigmáticos: el \textit{agujero negro}. De acuerdo con la teoría de la relatividad general de Einstein, un agujero negro puede definirse como una región del espacio-tiempo cuya curvatura es tan grande, que ninguna forma de materia o energía pueda escapar de su interior, ni siquiera la luz\footnote{En 1974 Stephen Hawking demostró teóricamente que al combinar la teoría cuántica de campos con la relatividad general, se encuentra que los agujeros negros deben emitir radiación térmica, lo que conduce a una evaporación gradual que culmina con una explosión de rayos gama y con la desaparición del agujero negro. Sin embargo, de momento no existe evidencia empírica concluyente en favor de las ideas de Hawking.} [6]. \\ 

En la época en que Chandrasekhar estableció su límite, las estrellas de neutrones y los agujeros negros no se conocían, y el destino final de un remanente estelar más masivo que una enana blanca era un completo misterio. Por lo tanto, no es de extrañar que el trabajo seminal de Chandrasekhar haya provocado desconcierto e incluso rechazo de parte de algunos de los más prominentes astrofísicos de aquel tiempo. Sin embargo, con el paso de los años las revolucionarias ideas del astrofísico indio comenzaron a ser aceptadas hasta convertirse en una parte fundamental del saber astronómico.

\section{La supernova champagne y el límite de Chandrasekhar}

Más de la mitad de las estrellas del universo están en la forma de \textit{sistemas binarios}, que son sistemas de dos estrellas que se encuentran tan próximas entre sí que están ligadas por su fuerza gravitatoria, orbitando alrededor de su centro de masas común. Las enanas blancas también se observan frecuentemente formando parte de sistemas binarios. \\ 

Cuando un sistema binario está compuesto por una enana blanca y una estrella normal, puede ocurrir que la enana blanca absorba material de la atmósfera de su compañera, alimentándose de ésta y aumentando gradualmente su masa. En el preciso momento en que la masa excede el límite de Chandrasekhar, el núcleo de la enana blanca alcanza la temperatura necesaria para provocar la fusión del carbono. En cuestión de segundos una fracción significativa de la materia que compone la enana blanca pasa por una reacción descontrolada, la cual liberara suficiente energía para provocar una colosal explosión denominada supernova de tipo Ia. Debido a que la explosión se produce siempre para un mismo valor de la masa, $1,4M_{\odot}$, la luminosidad generada es prácticamente la misma para cada supernova. La estabilidad de la luminosidad permite que estas explosiones sean usadas como estándares (cándelas estándar) para medir la distancia a las galaxias donde se producen las supernovas. Por tanto, la estabilidad de la luminosidad es también una evidencia en favor de la existencia del límite de Chandrasekhar. \\ 

El año 2003, investigadores de la Universidad de Toronto observaron una supernova tipo Ia cuya luminosidad era inusualmente alta, lo que parecía sugerir que la masa de la enana blanca progenitora superaba el límite de Chandrasekhar. De hecho, el valor estimado fue de aproximadamente $2M_{\odot}$ [19]. El astrónomo David Branch denominó a este evento supernova champagne [20] porque al parecer consideró que este fenómeno podía conducirnos a una nueva comprensión de las supernovas de tipo Ia, lo que a su juicio ameritaba una celebración descorchando una botella de champagne (el nombre oficial de la supernova es SN 2003fg). \\ 

¿Se violó realmente el límite de Chandrasekhar? La opinión generalizada parece ser que no se produjo una violación, y se han propuesto al menos dos posibles explicaciones de cómo una enana blanca pudo engordar tanto antes de convertirse en supernova. Una explicación es que la estrella original giraba tan rápido que la fuerza centrífuga evitaba que la gravedad la aplastara en el límite de Chandrasekhar. Otra posibilidad es que la explosión fue el resultado de la fusión de dos enanas blancas, lo que habría provocado un brillo superior al esperado [20]. El problema continúa abierto, y seguramente Chandrasekhar, que murió en 1995, habría estado muy entusiasmado con la noticia de la supernova champagne, pues como todo buen hombre de ciencia, amaba los rompecabezas y disfrutaba mucho intentado resolverlos.

\section{A modo de conclusión: El legado de Chandrasekhar}

Probablemente pocos acontecimientos científicos producen mayor admiración que el hecho que los astros obedezcan los designios de un simple mortal. Mediante un profundo dominio de la física de su tiempo, Subramayan Chandrasekhar fue capaz de prever un fenómeno astronómico que con el transcurso de los años ha sido confirmado por incontables observaciones astronómicas. El aporte científico de Chandrasekhar fue ampliamente reconocido cuando en el año 1983 la Academia de Ciencias de Suecia le otorgó el premio Nobel de Física “por sus estudios teóricos de los procesos físicos de importancia para la estructura y evolución de las estrellas”. \\ 

Después de su trabajo seminal sobre las enanas blancas, Chandrasekhar continuó desarrollando una extraordinaria y fructífera labor de investigación astronómica que abarcó temas tan diversos como el transporte radiativo en las estrellas, la estructura y la dinámica estelar, la teoría matemática de los agujeros negros, entre otros [21-24]. Por todo ello, Chandrasekhar es considerado justamente uno de los más grandes astrofísicos de nuestro tiempo.

\section*{Agradecimientos}
Quisiera gradecer al anónimo referee por su acuciosa revisión de este artículo y por sus útiles comentarios. 

\section*{Referencias}

[1] S. Chandrasekhar, The Astrophysical Journal, \textbf{74} (1931) 81-82.

\vspace{2mm}

[2] S. Chandrasekhar, Monthly Notices of the Royal Astronomical Society, \textbf{95} (1935) 207-225.

\vspace{2mm}

[3] K.L. Lang, Essential Astrophysics, Springer, Berlin, (2013).

\vspace{2mm}

[4] D. Maoz, Astrophysics in a Nutshell, 2 ed., Princeton University Press, Princeton, (2016).

\vspace{2mm}

[5] A.B. Bhattacharya, S. Joardar, R. Bhattacharya, Astronomy and Astrophysics, Infinity Science Press, New Delhi, (2008).

\vspace{2mm}

[6] M. Camenzind, Compact Objects in Astrophysics: White Dwarfs, Neutron Stars and Black Holes, Springer, Berlin, (2007).

\vspace{2mm}

[7] D. Gaefinkle, American Journal of Physics, \textbf{77} (2009) 683-687.

\vspace{2mm}

[8] C.B. Jackson, J. Taruna, S.L. Pouliot, B.W. Ellison, D.D. Lee, J. Piekarewicz, European Journal of Physics, \textbf{26} (2005) 695-712.

\vspace{2mm}

[9] F. Leblanc, An Introduction to Stellar Astrophysics, Wiley, Chichester, (2010).

\vspace{2mm}

[10] D. Maoz, Astrophysics in a Nutshell, Princeton University Press, Princeton, (2007).

\vspace{2mm}

[11] I. Sagert, M. Hempel, C. Greiner, J. Schaffner-Bielich, European Journal of Physics, \textbf{27} (2006).

\vspace{2mm}

[12] K. Krane, Modern Physics, 3 ed., John Wiley and Sons, Hoboken, (2012).

\vspace{2mm}

[13] P.A. Tipler, R.A. Llewellyn, Modern Physics, 6 ed., W. H. Freeman and Company, New York, (2012).

\vspace{2mm}

[14] J. Pinochet, M. Van Sint Jan, Physics Education, \textbf{51} 035007 (2016).

\vspace{2mm}

[15] R. Fowler, Monthly Notices of the Royal Astronomical Society, \textbf{87} (1926) 114-122.

\vspace{2mm}

[16] J.R. Oppenheimer, G.M. Volkoff,  Physical Review, \textbf{55} (1939) 374-381.

\vspace{2mm}

[17] R.C. Tolman, Physical Review, \textbf{55} (1939) 364-373.

\vspace{2mm}

[18] I. Bombaci, Astronomy and Astrophysics,\textbf{ 305} (1996) 871-877.

\vspace{2mm}

[19] D.A. Howell, Nature, \textbf{443} (2006) 308-311.

\vspace{2mm}

[20] D. Branch, Nature, \textbf{443} (2006) 283-284.

\vspace{2mm}

[21] S. Chandrasekhar, An Introduction to the Study of Stellar Structure, Dover, New York, (1958).

\vspace{2mm}

[22] S. Chandrasekhar, Radiative Transfer, Dover, New York, (1960).

\vspace{2mm}

[23] S. Chandrasekhar, The Mathematical Theory of Black Holes, Oxford University Press, New York, (1998).

\vspace{2mm}

[24] S. Chandrasekhar, Principles of Stellar Dynamics, Dover, New York (2005).

\end{document}